\documentclass[epj]{svjour}
\usepackage{graphics}
\usepackage[switch]{lineno}
\usepackage{amssymb,amsmath,amsfonts}
\usepackage{amsmath}
\usepackage{amsfonts}
\usepackage{amssymb}
\usepackage{graphicx}
\usepackage{color}

\input{tcilatex}
\begin{document}

\title{Rearrangement of electron shells and interchannel interaction \\
in the K photoabsorption of Ne}
\author{N. M. Novikovskiy\inst{1} \and D. V. Rezvan\inst{1} \and N. M. Ivanov%
\inst{1} \and I. D. Petrov\inst{2} \and B. M. Lagutin\inst{2} \\
A. Knie\inst{3} \and A. Ehresmann\inst{3} \and Ph. V. Demekhin\inst{3} \and %
V. L. Sukhorukov\inst{1}$^,$\inst{3}}
\offprints{novnim@mail.ru; vlsu16@mail.ru}
\institute{Institute of Physics, Southern Federal University, 344090 Rostov-on-Don,
Russia \and Rostov State Transport University, 344038 Rostov-on-Don, Russia %
\and  Institute of Physics and Center for Interdisciplinary Nanostructure
Science \\
and Technology (CINSaT), University of Kassel, D-34132, Kassel, Germany}
\date{Received: date / Revised version: date}


%

%

\authorrunning{N. M. Novikovskiy et al.}
\titlerunning{Rearrangement of
electron shells in K photoabsorption of Ne}

\abstract{
A detailed theoretical analysis of the $1s$ photoionization of neon is presented. It is found that the most significant many-electron correlation in computing photoionization of inner shells is the rearrangement of the outer shells caused by the inner vacancy. Further noticeable effects are: (i) the polarization of the ion core by the outgoing photoelectron and (ii) the coherent effect of double excitation/ionization. The core polarization increases the photoionization cross section by about 10\% at the 1s threshold, and the coherent excitation results in further increases by about 5\%. Incoherent excitation of the satellite channel leads to an additional 10\% increase in the photoabsorption cross section in the double-ionization threshold region.
\PACS{
{32.80.Fb}{Photoionization of atoms and ions}   \and
{32.80.Aa}{Inner-shell excitation and ionization}   \and
{31.15.V}{Electron correlation calculations for atoms, ions and molecules}}
} 
%

\maketitle%

\section{Introduction}

\label{sec:intro}

Recently, the interest in studying the photoabsorption by electrons of inner
atomic shells has been renewed as this process has been used to study the
interstellar medium (see, e.g., \cite{gatuzz15,muller17} and references
therein). Despite the fact that photoabsorption of inner atomic shells has
been studied for more than half a century, there are still noticeable
discrepancies between the various theoretical \cite%
{sukhorukov79,coreno99,kutzner99,gorczyca00} and experimental \cite%
{muller17,esteva83a,suzuki03} spectra even in case of the prototypical neon.

Photoabsorption cross section $\sigma _{1s}^{Abs}(\omega )$ of Ne as
function of the exciting-photon energy has been measured more than 50 years
ago \cite{liefeld65}. The experimental resolution in that work was
sufficient to observe spectral features associated with Rydberg series
whereas the $\sigma _{1s}^{Abs}(\omega )$ cross section has been determined
on an arbitrary scale. The absolute photoionization cross section for the $%
1s $ shell of Ne has been computed in \cite{sukhorukov79} considering the
rearrangement of atomic shells by the $1s$ vacancy. Using the synchrotron
radiation allowed Esteva et al. \cite{esteva83a} to obtain cross section $%
\sigma _{1s}^{Abs}(\omega )$ on an absolute scale and to reveal some
additional above-threshold sharp spectral features
which have been interpreted as shake satellites connected mainly with the $1s^{-1}2p^{-1}3p^{2}~^{1}P$ terms. The cross
sections for production of those satellites have been calculated in our
earlier paper \cite{sukhorukov87} with taking into account the
doubly-excited states. It was revealed that the satellites observed in \cite{esteva83a} in addition to the $1s^{-1}2p^{-1}3p^{2}~^{1}P$ terms are also determined by the states of the $1s^{-1}2p^{-1}3p4p$ configurations and by the strong many-electron correlation known as the dipole polarization of electron shells (DPES) \cite{sukhorukov18} which is
described mainly by the $3p3p-3s(n/\varepsilon)(s/d)$ excitations.
In paper \cite{sukhorukov87} it was also mentioned that an interaction between the autoionizing doubly-excited states and respective continua can result in Fano-type [12] resonance profiles. These resonances are energetically located starting from about 30 eV above the $1s$- threshold. Therefore, the precise theoretical description of those resonances can be separated from the problem of computing the smooth cross section
in the energy region just above the $1s$ threshold.

Kutzner and Rose \cite{kutzner99} modified the relativistic random-phase
approximation to include the rearrangement of the ionic core and computed
the Ne $\sigma_{1s}^{Abs}(\omega)$ cross section adding the background from
the photoionization of the $2s$ and $2p$ shells. Gorczyca \cite{gorczyca00}
applied the R-matrix MQDT optical potential method in order to compute the
photoabsorption cross section in the region of the $1s^{-1}np$ resonances
and just above the $1s$ threshold.

An absolute Ne $\sigma_{1s}^{Abs}(\omega)$ cross section has been determined
anew by Suzuki and Saito \cite{suzuki03} using monochromatized synchrotron
radiation in the extended energy range between 50--1300 eV, but with broad
bandwidth. The absolute threshold photoabsorption cross section $\sigma_{1s}^{Abs}(\omega_{thr})$ was about 10\% smaller than previously measured by Esteva et al. \cite{esteva83a}. Near-threshold photoabsorption cross section $\sigma_{1s}^{Abs}(\omega)$ of Ne has been measured by Prince et al. \cite{prince05} with high resolution. This experiment revealed that the
doubly-excited shake resonances have the asymmetric Fano shape as predicted
in \cite{sukhorukov87} but resulted in the threshold cross section $\sigma
_{1s}^{Abs}(\omega_{thr})=0.45$ Mb which is substantially larger than that
measured by Suzuki and Saito \cite{suzuki03} $\sigma_{1s}^{Abs}(\omega
_{thr})=0.376$ Mb. The reason of this observation is not completely clear
but supposedly it can be connected with overestimated photoabsorption by the $2s$
and $2p$ shells obtained in \cite{prince05}. Indeed, the value of this cross
section $\sigma_{2s+2p}(\omega_{thr})=0.09$ Mb measured in \cite{prince05}
is about 4 times larger than that reported in \cite{suzuki03} $\sigma
_{2s+2p}(\omega_{thr})=0.0228$ Mb. High-resolution near-threshold $1s$
photoabsorption cross section has been determined by M\"{u}ller et al. \cite%
{muller17}. Both experiments \cite{muller17,suzuki03} yield cross sections
which are smaller by about 10\% in the threshold region than that measured
previously \cite{esteva83a}. A similar difference exists between the
calculations \cite{sukhorukov79,gorczyca00} and \cite{kutzner99}.

In the present study, we calculate the Ne $\sigma_{1s}^{Abs}(\omega)$ cross
section with `step-by-step' inclusion of different many-electron
correlations concentrating on the `smooth' region which is free of the
doubly-excited resonances. This procedure allows to clarify the individual
role of each many-electron correlation in the photoionization of inner
shells.

\section{Method of calculation}

\label{sec:method}

Photoabsorption of Ne near the $1s$ threshold is determined mainly by the
channels shown in scheme (\ref{eq:scheme}). Background absorption stems from
the $2s\dashrightarrow\varepsilon p$\ and $2p\dashrightarrow\varepsilon
\{s,d\}$\ photoionization. The latter processes produce photoelectrons with
large energies at the $1s$ threshold and, therefore, are not strongly
affected by many-electron correlations in this exciting-photon energy range.
Contrary, processes (\ref{eq:scheme}) release photoelectrons with small
energies, which results in a substantial influence of many-electron
correlations on the cross sections of channels (\ref{eq:scheme}main) and (%
\ref{eq:scheme}sat):

\begin{equation}
\begin{array}{cccc}
& \dashrightarrow & 1s^{1}2s^{2}2p^{6}\left( n/\varepsilon\right) p &
\mathrm{(main)} \\
1s^{2}2s^{2}2p^{6} &  & \mathrm{main~line} &  \\
&  & \updownarrow &  \\
& \dashrightarrow & 1s^{1}2s^{2}2p^{5}\left( n/\varepsilon\right) ^{\prime
}p\ \left( n/\varepsilon\right) ^{\prime\prime}p & ~~\mathrm{(sat)} \\
&  & \mathrm{satellite} &
\end{array}
\label{eq:scheme}
\end{equation}

Transition amplitudes are composed of the direct $\mathbf{D}$ and
correlational $\mathbf{D}^{\mathrm{corr}}$ parts:

\begin{equation}
A_{1s}^{\varepsilon p}=\left\langle 1s^{1}2s^{2}2p^{6}\varepsilon
p\left\Vert \mathbf{D}+\mathbf{D}^{\mathrm{corr}}\right\Vert
1s^{2}2s^{2}2p^{6}\right\rangle  \label{eq:amp-tot}
\end{equation}
where

\begin{equation}
\langle1s^{1}2s^{2}2p^{6}\varepsilon p||\mathbf{D}||1s^{2}2s^{2}2p^{6}%
\rangle=\sqrt{2}\langle\varepsilon p\left\vert \mathbf{d}\right\vert
1s\rangle  \label{eq:amp-dir}
\end{equation}
and

\begin{multline}
\left\langle 1s^{1}2s^{2}2p^{6}\varepsilon p\left\Vert \mathbf{D}^{\mathrm{%
corr}}\right\Vert 1s^{2}2s^{2}2p^{6}\right\rangle =\sqrt{2}\times
\label{eq:amp-corr} \\
\sum_{\varepsilon^{\prime},\varepsilon^{\prime\prime}}\frac{%
(6F^{0}-G^{0}-0.4G^{2})\langle\varepsilon^{\prime}p|2p\rangle\langle%
\varepsilon ^{\prime\prime}p\left\vert \mathbf{d}\right\vert 1s\rangle}{%
IP(1s^{-1})+\varepsilon-IP(1s^{-1}2p^{-1})-\varepsilon^{\prime}-\varepsilon
^{\prime\prime}+\imath\delta}~.
\end{multline}
In equation~(\ref{eq:amp-corr}), $IP$ denotes the ionization potential of
the respective state; the sum over $\varepsilon^{\prime}$ and $\varepsilon
^{\prime\prime}$ includes summation over the discrete and integration over
the continuum states; the notations $F^{0}\equiv R^{0}(2p\varepsilon
p;\varepsilon^{\prime}p\varepsilon^{\prime\prime}p)$ and $G^{k}\equiv
R^{k}(2p\varepsilon p;\varepsilon^{\prime\prime}p\varepsilon^{\prime}p)$ for
the Slater integrals are used for brevity.

The one-electron matrix elements of the electric dipole transition operator,
$\mathbf{d}$, entering equations (\ref{eq:amp-dir}), (\ref{eq:amp-corr}) are
given by \cite{sukhorukov79,sukhorukov87}:

\begin{multline}
\langle\varepsilon p\left\vert \mathbf{d}\right\vert 1s\rangle=\left\langle
1s^{+}|1s\right\rangle \left\langle 2s^{+}|2s\right\rangle ^{2}\left\langle
2p^{+}|2p\right\rangle ^{6}  \label{eq:amp1sep} \\
\times\left[ \left\langle \varepsilon p^{+}\left\vert d\right\vert
1s\right\rangle -\left\langle 2p^{+}\left\vert d\right\vert 1s\right\rangle
\frac{\left\langle \varepsilon p^{+}|2p\right\rangle }{\left\langle
2p^{+}|2p\right\rangle }\right. \\
-\left. \left\langle \varepsilon p^{+}\left\vert d\right\vert
2s\right\rangle \frac{\left\langle 2s^{+}|1s\right\rangle }{\left\langle
2s^{+}|2s\right\rangle }\right] ,
\end{multline}
where $\left\langle \mathrm{bra}^{+}\right\vert $ and $\left\vert \mathrm{ket%
}\right\rangle $ AOs are computed in configurations $1s^{1} 2s^{2} 2p^{6}$
and $1s^{2} 2s^{2} 2p^{6}$, respectively; $\left\langle
n^{\prime}\ell^{\prime}\left\vert d\right\vert n\ell\right\rangle $ is the
radial integral computed either in length or in velocity gauges, depending
on the sign in equation (\ref{eq:sig-1sep}). The transition amplitude (\ref%
{eq:amp-tot}) determines the photoionization cross section of the $%
1s\dashrightarrow\varepsilon p$\ transition as:

\begin{equation}
\sigma_{1s}(\omega)=\frac{4}{3}\pi^{2}\alpha
a_{0}^{2}\omega^{\pm1}\left\vert A_{1s}^{\varepsilon p}\right\vert ^{2},
\label{eq:sig-1sep}
\end{equation}
where the signs ($+$) and ($-$) correspond to the length and velocity forms
of operator $\mathbf{d}$, respectively; $\omega$ denotes the exciting photon
energy in atomic units; $\alpha=1/137.036$ is the fine structure constant;
the square of the Bohr radius $a_{0}^{2}=28.0028$ Mb converts atomic units
for cross sections to $\mathrm{Mb}=10^{-22}~\mathrm{m}^{2}$. The exciting
photon energy $\omega$ and the photoelectron energy, $\varepsilon$, counted
from the $1s$ threshold, $IP(1s^{-1})$, are related via:

\begin{equation}
\omega=IP(1s^{-1})+\varepsilon.  \label{eq:Ephot}
\end{equation}

\begin{figure}[ptb]
\resizebox{0.48\textwidth}{!}{\includegraphics{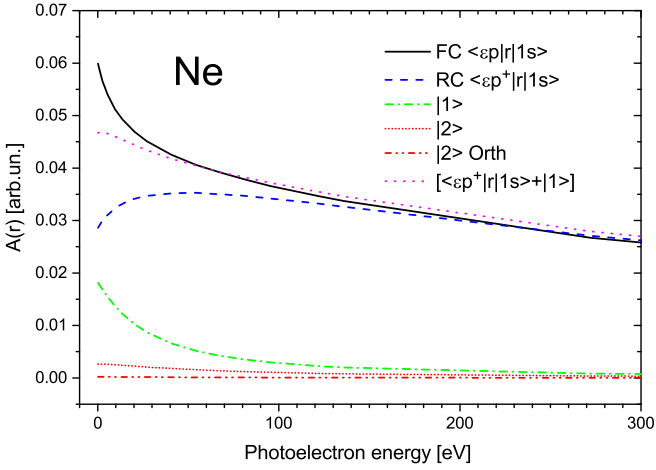}}
\caption{Comparison between partial transition amplitudes computed within
the frozen core approach (FC), different terms contained in square brackets
of equation (\protect\ref{eq:amp1sep}), and the sum of these terms.
Dash-double dotted and dotted curves show the 3rd term in square brackets of
(\protect\ref{eq:amp1sep}) computed with and without taking into account
orthogonality condition, respectively (see text).}
\label{fig:Fig_ampl}
\end{figure}

The values of terms in square brackets for the transition amplitude (\ref%
{eq:amp1sep}) are depicted as functions of electron energy in Fig.~\ref%
{fig:Fig_ampl}. The transition amplitude computed in the frozen core
approximation FC is shown in the same figure for comparison. One can
recognize that the relaxation of the ionic core strongly decreases the
amplitude $\left\langle \varepsilon p\left\vert d\right\vert 1s\right\rangle
$ computed within the FC approach whereas additional terms compensate this
decrease at $\varepsilon\leq150$ eV. It is important to note that the
application of the theory of non-orthogonal orbitals implies the
orthogonality of the total wave function of an ionic state to the all
low-lying state wave functions. In the considered case of Ne it means that
the total wave function for the configuration $1s^{1}2s^{2}2p^{6}$ should be
orthogonal to the $1s^{2}2s^{1}2p^{6}$ one. The fulfilment of this
requirement results in a decrease of the 3rd term in square brackets of
equation (\ref{eq:amp1sep}) by about one order of magnitude (cf. short
dotted and dash-double dotted curves in Fig.~\ref{fig:Fig_ampl}), allowing
to neglect these terms in the calculations.

The photoionization cross section for the satellite production (pathway (\ref%
{eq:scheme}sat)) was calculated via the following equation adopted from
Refs. \cite{sukhorukov87,sukhorukov91a} for the case of the $%
1s2p\dashrightarrow(n/\varepsilon)^{\prime}p(n/\varepsilon)^{\prime\prime}p$%
\ transition in Ne:

\begin{multline}
\sigma_{1s2p}(\omega)=\frac{4\pi^{2}\alpha a_{0}^{2}}{3}
\label{eq:sig-1sep2} \\
\frac{12\ \omega}{1+\delta_{n^{\prime}n^{\prime\prime}}}\int_{0}^{%
\omega-IP(1s^{-1}2p^{-1})}\left[ A^{2}+B^{2}-AB/3\right] d\varepsilon
^{\prime\prime},
\end{multline}
where

\begin{align}
A & =\left\langle (n/\varepsilon)^{\prime}p\left\vert d\right\vert
1s\right\rangle \left\langle (n/\varepsilon)^{\prime\prime}p|2p\right\rangle
,  \label{eq:amp-double} \\
B & =\left\langle (n/\varepsilon)^{\prime\prime}p\left\vert d\right\vert
1s\right\rangle \left\langle (n/\varepsilon)^{\prime}p|2p\right\rangle ,
\notag
\end{align}
and integration over $\varepsilon^{\prime\prime}$\ is carried out over the
following surface:

\begin{equation}
\omega=IP(1s^{-1}2p^{-1})+\varepsilon^{\prime}+\varepsilon^{\prime\prime
},~~~~\varepsilon^{\prime}\leq\varepsilon^{\prime\prime}.  \label{eq:surface}
\end{equation}
If either photoelectron or excited electron belongs to a discrete $n^{\prime
}p$ state, the integration over the $\varepsilon^{\prime\prime}$ in equation
(\ref{eq:sig-1sep2}) is lifted, and $\varepsilon^{\prime\prime}$ is
determined as:

\begin{equation}
\omega=IP(1s^{-1}2p^{-1}n^{\prime}p)+\varepsilon^{\prime\prime}.
\label{eq:point}
\end{equation}

\section{Results and discussion}

\label{sec:results}

In order to clarify the influence of different many-electron correlations on
the the inner-shell photoionization we computed cross section $\sigma
_{1s}^{\varepsilon p}(\omega)$ of Ne in different approximations:

\begin{description}
\item[\textbf{FC}] Frozen core approach, where the relaxation of atomic
orbitals (AOs) by the $1s$ vacancy was neglected. The core orbitals were
obtained in the neutral ground configuration $1s^{2}2s^{2}2p^{6}$, and were
also used to compute the AO of the photoelectron in the $1s^{1}2s^{2}2p^{6}%
\varepsilon p~^{1}P$ state.

\item[\textbf{RC}] Relaxed core approach, which takes into account the
relaxation of AOs. The core AOs of the initial state were taken from the FC
approach; for the final state core AOs $P_{n\ell}^{+}(r)$ were computed in
the self-consistent configuration $1s^{1}2s^{2}2p^{6}$; the photoelectron
AOs $P_{\varepsilon p}^{+}(r)$ were computed in the `relaxed' core AOs.

\item[\textbf{RAC}] Rearranged core approach, where additional second term
enters the transition amplitude (\ref{eq:amp1sep}).

\item[\textbf{RAC+CP}] \emph{Ab initio} core polarization potential \cite%
{petrov99} was included into the equation for the photoelectron AOs $%
P_{\varepsilon p}^{+}(r)$ in addition to the RAC approach.

\item[\textbf{RAC+CP+IC}] The coherent correlational part of the transition
amplitude $\mathbf{D}^{\mathrm{corr}}$\ (\ref{eq:amp-corr}) caused by the
interaction between the `satellite' $1s2p-\varepsilon^{\prime}p\varepsilon
^{\prime\prime}p$ and the `main' $1s-\varepsilon p$ channels was taken into
account in addition to the approximations discussed above.
\end{description}

\begin{figure}[ptb]
\resizebox{0.48\textwidth}{!}{\includegraphics{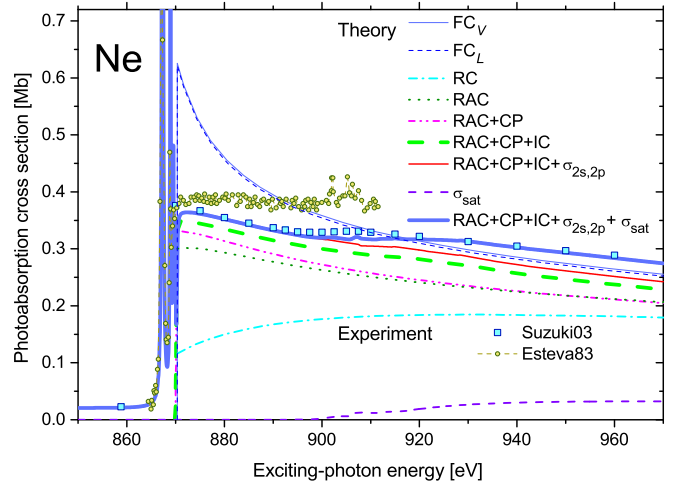}}
\caption{Photoionization cross sections $\protect\sigma_{1s}(\protect\omega)$
computed in the approximations described in the text. The lowest dashed
curve shows the cross section for the satellite production $\protect\sigma%
_{1s2p}(\protect\omega)$. The thick solid line represents the computed
photoabsorption cross section $\protect\sigma_{1s}^{Abs}(\protect\omega)$.
Experimental photoabsorption data `Esteva83' and `Suzuki03' are from Refs.
\protect\cite{esteva83a} and \protect\cite{suzuki03}, respectively. The size
of points representing experiment \protect\cite{suzuki03} corresponds to
error bars claimed in this work.}
\label{fig:Fig_result}
\end{figure}

Photoionization cross sections $\sigma_{1s}(\omega)$ computed in the
described approximations are depicted in Fig.~\ref{fig:Fig_result}. Velocity
gauge is used for presentation, because length and velocity results agree
fairly well (cf. cross sections FC$_{L}$ and FC$_{V}$ computed in the length
and velocity gauges, respectively). In order to simplify the figure, the
below-threshold region is shown only for the final result (thick solid line).

In Fig.~\ref{fig:Fig_result}, we set the computed $1s$ threshold to the
measured experimental value $IP_{Exp}(1s^{-1})=870.33$ eV, which was
estimated by us using the Rydberg formula and the data of Ref. \cite%
{muller17}. The estimated value agrees with the value recommended by NIST, $%
IP_{Exp}(1s^{-1})=870.23(18)$ eV \cite{deslattes14}, within the error bar.
In the present work, the value of the $1s$ threshold was computed within the
Pauli-Fock approximation, which uses the Breit operator for description of
the relativistic corrections \cite{kau97}. The computed value of $%
IP_{PF}(1s^{-1})=869.55$ eV, is by 0.78 eV smaller then the measured one
because of neglecting high-order corrections of perturbation theory to the
total energies of the $1s^{2}2s^{2}2p^{6}$ and $1s^{1}2s^{2}2p^{6}$
configurations.

From Fig.~\ref{fig:Fig_result}, one can see that the frozen core
approximation FC overestimates the cross section at the $1s$ threshold $%
\sigma_{1s}(\omega_{thr})$. Taking into account the core relaxation (RC
approach) decreases $\sigma_{1s}(\omega_{thr})$ by a factor of 5.6. The
additional term in equation (\ref{eq:amp1sep}) (RAC approach) increases $%
\sigma_{1s}(\omega_{thr})$ by a factor of 2.7, thus partly compensating the
previous decrease. The further increase of the $\sigma_{1s}(\omega_{thr})$
by about 10\% stems from taking into account the core polarization
potential. The increase of $\sigma_{1s}(\omega_{thr})$ due to the coherent
pathway (\ref{eq:scheme}sat) is twice smaller than due to core polarization
but results in the appearance of two inflection points in the $%
\sigma_{1s}(\omega)$ curve at $\omega=913.0$ eV and $\omega=934.5$ eV. It is
very desirable to confirm the existence of these points experimentally
applying, e.g., photoelectron spectroscopy.

In order to obtain the photoabsorption cross section $\sigma_{1s}^{Abs}(%
\omega)$, the photoionization cross sections of the outer shells $\sigma
_{2s}(\omega)$, $\sigma_{2p}(\omega)$ and the cross section for the
satellites production $\sigma_{1s2p}(\omega)$ (\ref{eq:sig-1sep2}) were
computed. It turned out that intrashell and intershell correlations \cite%
{amusia90a} have small influence on $\sigma_{2s}(\omega)$ and $%
\sigma_{2p}(\omega)$ because of the large photoelectron energy $\varepsilon$
at the $1s$ threshold. The absolute values of $\sigma_{2s}(%
\omega_{thr})=0.014$ Mb and $\sigma _{2p}(\omega_{thr})=0.005$ Mb at the $1s$
threshold are small compared to the $\sigma_{1s}(\omega_{thr})=0.364$ Mb.
The resulting value of $\sigma _{2s+2p}(\omega_{thr})=0.019$ Mb is slightly
less than the background value of $\sigma_{2s+2p}(\omega_{thr})=0.0228$ Mb
determined experimentally \cite{suzuki03}.

The cross section for the satellites production $\sigma_{1s2p}(\omega)$
computed according to equation (\ref{eq:sig-1sep2}) is presented by the
lowest dashed curve in Fig.~\ref{fig:Fig_result}.
Precise calculation of the
lineshapes produced by the autoionizing doubly-excited $1s^{1}2p^{5}n\ell
n^{\prime}\ell^{\prime}$ resonances is a separate cumbersome problem (see,
e.g. \cite{sukhorukov12}) which is out of scope of the present paper.
Therefore, we computed the integral cross section for the production of the $1s2p-npn^{\prime}p$ satellites via equation (\ref{eq:sig-1sep2}) and
presented them by Gaussians with FWHM of 1.6 eV located at the center of
gravity of the respective $1s^{1}2p^{5}npn^{\prime}p$ configuration. This
Gaussian FWHM was obtained using the transition array technique \cite{bauche-arnoult79,bauche-arnoult82}. This technique simulates the multiplet
splitting of the configurations with many open shells and it is known to be
very effective in the overall description of complicated atomic spectra
\cite{bauche88a}.

\begin{figure}[ptb]
\resizebox{0.48\textwidth}{!}{\includegraphics{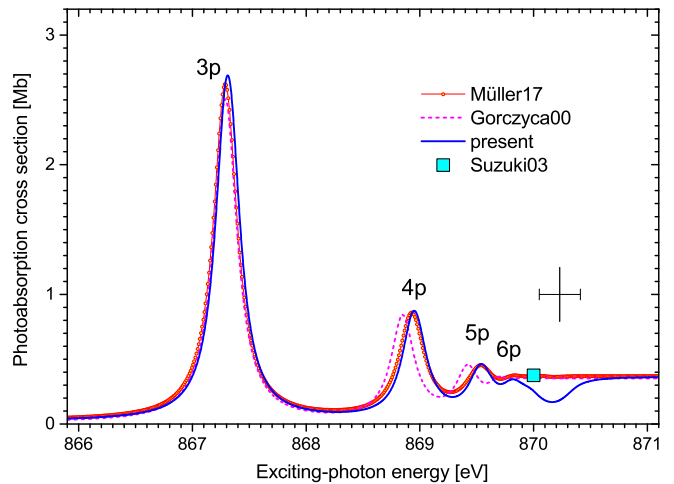}}
\caption{Comparison between the photoabsorption cross section $\protect\sigma%
_{1s}^{Abs}(\protect\omega)$ computed in the present work, in the work of
Gorczyca \protect\cite{gorczyca00} and experimentally determined by M\"{u}%
ller et al. \protect\cite{muller17}. A single experimental point obtained by
Suzuki and Saito \protect\cite{suzuki03} falling in the depicted energy
interval is also shown. The energy scale of Gorczyca \protect\cite%
{gorczyca00} was set to the experimental scale of Ref. \protect\cite%
{muller17} using the first peak related with the $1s\dashrightarrow3p$
transition, whereas our energy scale was set to the experimental one using
the $1s$ threshold (see text).}
\label{fig:Fig_edge}
\end{figure}

In Fig.~\ref{fig:Fig_result}, the resulting photoabsorption cross section is
denoted as `\textbf{RAC +CP +IC +}$\sigma_{2s,2p}(\omega)$ +$\sigma _{%
\mathrm{sat}}(\omega)$'. One can see that the presently computed cross
section $\sigma_{1s}^{Abs}(\omega)$ is in excellent agreement with the data
of Suzuki and Saito \cite{suzuki03}, whereas it differs from $%
\sigma_{1s}^{Abs}(\omega)$ measured by Esteva et al. \cite{esteva83a}. It
could be of great interest to clarify the reason of this disagreement.

The computed below-threshold photoabsorption cross sections are presented in
an enlarged scale in Fig.~\ref{fig:Fig_edge} by the thick solid curve. The
minimum in presently computed $\sigma_{1s}^{Abs}(\omega)$ at $\omega$ about
870.2 eV stems from considering the $1s^{-1}np$ $(n=3-7)$ resonances only,
whereas the experiment includes the infinite $np$ Rydberg series. Lines
corresponding to the $1s-np$\ transitions are convolved with Lorentzians of
FWHM 0.230 eV and with their area computed as:

\begin{multline}
f_{\sigma}\left( 1s-np\right) \left[ \mathrm{Mb\cdot eV}\right] =
\label{eq:gf} \\
2\pi^{2}\alpha a_{0}^{2}\cdot f\left( 1s-np\right) \cdot2\mathrm{Ry[eV]},
\end{multline}
where $f$ is the dimensionless oscillator strength of the $1s-np$ transition
converted to the area under the $\sigma_{1s}(\omega)$ curve (or
`dimensioned' oscillator strength $f_{\sigma}$ which has a meaning of the
photoionization cross section) by multiplying with a factor $2\pi^{2}\alpha
a_{0}^{2}\cdot2\mathrm{Ry}$. The FWHM was taken equal to the natural width
of the $1s$\ level $\Gamma_{1s}$, which was computed in the present work
using the self-consistent AOs of the $1s^{1}2s^{2}2p^{6}$ configuration.
This approximation is known to result in adequate values of natural widths
\cite{howat78a}. $\Gamma_{1s}=0.230$ eV agrees well with $\Gamma_{1s}=0.241$
eV computed in \cite{gorczyca00} and is slightly smaller than the
experimental value $\Gamma_{1s}=0.271(20)$ eV \cite{coreno99}.

In Fig.~\ref{fig:Fig_edge}, the $\sigma_{1s}^{Abs}(\omega)$ computed below
the ionization threshold is compared with theoretical data of Gorczyca \cite%
{gorczyca00} and experimental data of M\"{u}ller et al. \cite{muller17} in
an enlarged scale. Small differences between energy positions of the $%
1s^{-1}np$ resonances and their oscillator strengths $f_{\sigma}$ computed
in \cite{gorczyca00} and measured in \cite{muller17} could be connected with
neglecting the core polarization in those calculations \cite{gorczyca00}. An
excellent agreement between our calculation and experiment \cite{muller17}
is evident from Fig~\ref{fig:Fig_edge}.

\begin{table}[ptb]
\caption{Oscillator strengths $f_{\protect\sigma}\left( 1s-np\right) $ (%
\protect\ref{eq:gf}) computed for Ne in different approximations.}
\label{tab:par1s-np}%
\begin{tabular}{rrrrrrr}
\hline\hline
$np$ & \multicolumn{5}{c}{Approximation} & Exp. \\ \cline{2-6}
& FC & RC & RAC & +CP$^{a}$ & +IC$^{b}$ & \cite{muller17}$^{c}$ \\
&  &  &  &  &  &  \\ \hline
$3p$ & 3.513 & 0.281 & 0.827 & 0.964 & 1.010 & 1.027 \\
$4p$ & 0.733 & 0.094 & 0.258 & 0.294 & 0.308 & 0.307 \\
$5p$ & 0.287 & 0.042 & 0.113 & 0.127 & 0.134 & ~ \\ \hline\hline
\end{tabular}
\newline
\newline
$^{a}$ Approximation \textbf{RAC+CP} (see text) \newline
$^{b}$ Approximation \textbf{RAC+CP+IC} \newline
$^{c}$ Values are estimated by us (see text)
\end{table}

Oscillator strengths $f_{\sigma}\left( 1s-np\right) $\ (\ref{eq:gf})
computed in the different approximations are listed in Table~\ref%
{tab:par1s-np} in comparison to experimental data. Experimental values were
obtained by fitting cross sections with a constant background (appeared to
be 0.025(1) Mb) and Lorentzians. One can see that the general trends in the
computed cross sections between different approximations is similar to that
for the photon energies above the ionization threshold, but they become more
pronounced. For instance, the relaxation of the core under the $1s$ vacancy
resulted in a decrease of $f_{\sigma}\left( 1s-3p\right) $ by a factor of $%
12.5$ with subsequent compensation due to additional terms by a factor of $%
2.9$. Taking into account of the core polarization increases $%
f_{\sigma}\left( 1s-3p\right) $ by about 17\%. An excellent agreement
between the final result obtained in the present work and the experimental
value from \cite{muller17} can be seen from the last two columns of Table~%
\ref{tab:par1s-np}.

\section{Conclusions}

\label{sec:fin}

The absolute photoabsorption cross sections as a function of the
exciting-photon energy $\sigma_{1s}^{Abs}(\omega)$ was calculated for Ne
with taking into account important many electron correlation effects. To
solve this problem we computed photoionization cross sections $%
\sigma_{1s}(\omega)$, $\sigma_{2s}(\omega)$, $\sigma_{2p}(\omega)$\ and the
cross section $\sigma_{1s2p}(\omega)$ for the satellite production. The
influence of different many-electron correlations on the computed $%
\sigma_{1s}(\omega)$ was investigated in details. The following
many-electron correlation are important in describing the near-threshold
photoabsorption of the Ne $1s$ shell (in decreasing order): (i) relaxation
of the core atomic orbitals caused by the $1s$ vacancy; (ii) rearrangement
of the transition amplitude caused by the relaxation of atomic orbitals;
(iii) polarization of the core by the outgoing (or excited) electron; (iv)
coherent interaction between main and satellite photoionization channels.
Excellent agreement between the present theory and recent experiment allows
to conclude the adequacy of the approximation used here for the description
of the inner-shell photoionization. Coherent interaction between the main
and satellite channels results in the appearance of two inflection points in
the $\sigma_{1s}(\omega)$ photoionization function in the region of the
double-ionization threshold. It could be interesting to check this
prediction experimentally.


\section{Authors contributions}

VLS conceived the idea and coordinated the research pro-ject; PhVD designed,
produced and tested computer programs; NMN, IDP, DVR, NMI performed the
calculations and combined the results; BML, AK, AE participated in
discussion of results and contributed to writing manuscript. All authors
were involved in the preparation of the manuscript, commented and approved
its final version.

\begin{acknowledgement}
This work was partly supported by the Deutsche Forschungsgemeinschaft (DFG) within the Sonderforschungsbereich SFB--1319 `Extreme Light for Sensing and Driving Molecular Chirality -- ELCH'. NMN, IDP, BML and VLS would like to thank the Institute of Physics, University of Kassel for the hospitality. VLS appreciate support from Southern Federal University within the inner project no~3.6105.2017/BP. IDP and BML appreciate support from the grant no~16-02-00771a of the RFBR.
We are grateful to Prof. A. M\"uller and Prof. T.W. Gorczyca for providing in digital form data from papers \cite{muller17} and \cite{gorczyca00}, respectively.
\end{acknowledgement}


\end{document}